\documentclass[conference]{IEEEtran}
\IEEEoverridecommandlockouts
\usepackage{amsmath,amssymb,amsfonts}
\usepackage{algorithmic}
\usepackage{graphicx}
\usepackage{textcomp}
\usepackage{xcolor}
\usepackage{array}
\usepackage{multirow}
\usepackage{threeparttable}
\usepackage[sorting=none]{biblatex}
\newcolumntype{C}[1]{>{\centering\arraybackslash}m{#1}}
\def\BibTeX{{\rm B\kern-.05em{\sc i\kern-.025em b}\kern-.08em
    T\kern-.1667em\lower.7ex\hbox{E}\kern-.125emX}}
\begin{document}

\title{HuntGPT: Integrating Machine Learning-Based Anomaly Detection and Explainable AI  with Large Language Models (LLMs)\\}
\author{\IEEEauthorblockN{Tarek Ali and Panos Kostakos}
\IEEEauthorblockA{\textit{Faculty of Information Technology and Electrical Engineering}\\
\textit{Center for Ubiquitous Computing}\\
\textit{University of Oulu}\\
Oulu, Finland 90570 \\
Tarek.Ali, Panos.Kostakos@oulu.fi}
}

\maketitle

\begin{abstract}

Machine learning (ML) methods for network anomaly detection are emerging as effective proactive strategies in threat hunting, substantially reducing the time required for threat detection and response. However, the challenges in training and maintaining ML models, coupled with frequent false positives, diminish their acceptance and  trustworthiness.  In response, Explainable AI (XAI) techniques have been introduced to enable cybersecurity operations teams to assess alerts generated by AI systems more confidently. Despite these advancements, XAI tools have encountered limited acceptance from incident responders and have struggled to meet the decision-making needs of both analysts and model maintainers. Large Language Models (LLMs) offer a unique approach to tackling these challenges. Through tuning, LLMs have the ability to discern patterns across vast amounts of information and meet varying functional requirements. In this research, we introduce the development of HuntGPT, a specialized intrusion detection dashboard created to implement a Random Forest classifier trained utilizing the KDD99 dataset. The tool incorporates XAI frameworks like SHAP and Lime, enhancing user-friendliness and intuitiveness of the model. When combined with a GPT-3.5 Turbo conversational agent, HuntGPT aims to deliver detected threats in an easily explainable format, emphasizing user understanding and offering a smooth interactive experience. We investigate the system’s comprehensive architecture and its diverse components, assess the prototype's technical accuracy using the  Certified Information Security Manager (CISM) Practice Exams, and analyze the quality of response readability across six unique metrics. Our results indicate that conversational agents, underpinned by LLM technology and integrated with XAI, can enable a robust mechanism for generating explainable and actionable AI solutions, especially within the realm of intrusion detection systems.

\end{abstract}


\section{Introduction}

In recent decades, there has been a substantial escalation in cyber-attacks targeting critical and enterprise infrastructure. By 2025, anticipated annual financial damages from these cyber-attacks are projected to reach \$10.5 trillion USD, a substantial leap from \$3 trillion USD recorded in 2015 \cite{cybersecurityventures2023Cybersecurity}. To counteract the evolving cyber threats, the National Institute of Standards and Technology (NIST) introduced a Cybersecurity Framework in 2014. This framework prescribes iterative cybersecurity policies for identification, protection, detection, response, and recovery processes related to cyber incidents \cite{cybersecurity2018framework}.

In this backdrop, human experts play a vital role in analysing extensive telemetry data and Indicators of Compromise (IoC) to isolate real threats \cite{shu2018threat, NAD-survey}. Consequently, building on the foundation laid by the NIST Framework, an extensive ecosystem—comprising tools, methodologies, and techniques—has been established to enable the proactive identification of threats, a process referred to as Cyber Threat Hunting (CTH) \cite{101145ww, 9672347, 9791234}. Threat hunting tools enable analysts to apply their specialized knowledge to formulate and test threat hypotheses by analysing system telemetry as well as threat intelligence from external sources \cite{shu2018threat}. 

Machine learning-based anomaly detection tools are particularly noteworthy, designed to uncover both known and unknown threats. Generally, network anomalies are categorized into performance-related, such as file server failures and transient congestion, and security-related anomalies like Denial of Service attacks, spoofing, and network intrusions \cite{anomaly-type}. Evidently, the infusion of machine learning into CTH tools has notably increased the incidence of false positives in real-world operational environments \cite{cavallaro2023machine}.

Explainable Artificial Intelligence (XAI) is at the focus of several proposed conceptual enhancements to existing cybersecurity frameworks, aiming to address the challenges brought forth by integrations of machine learning \cite{charmet2022explainable}. A pivotal advancement in this realm is the evolution of \textit{Cybertrust} frameworks, accentuating the necessity to integrate explainable, interpretable, and actionable AI in cybersecurity operations \cite{9187463}. Nevertheless, the swift advancements in the domain may lead to information overload for incident responders and ML model maintainers, potentially resulting in a sluggish adoption rate \cite{nyre2022explainable}. 

Large Language Models (LLMs), driving the rapid development of autonomous agents, show significant potential to transform the landscape of cybersecurity. Their capability to seamlessly integrate diverse AI tasks and adapt to various use cases, positions them as versatile solutions that could boost the adoption of XAI as well as drive down operational costs. Specifically, Large Language Models, and conversational agents in particular, have showcased outstanding capabilities in promoting applications for actionable AI, which are vital in providing response suggestions to threat responders.

In this paper, we introduce a novel prototype, HuntGPT, aimed at integrating actionable, interpretable, and explainable AI in cybersecurity operations. The prototype is designed to perform analysis on network traffic, utilizing a Random Forest classifier as the anomaly detection model. This model, trained utilizing the KDD99 dataset \cite{kdd99}, is deployed to classify the acquired packets systematically. We utilize explainability frameworks like SHAP and LIME, in tandem with a conversational agent powered by the Language Learning Model API from OpenAI, predominantly known to users as ChatGPT. The prototype undergoes evaluation for technical accuracy through the utilization of Certified Information Security Manager Practice (CISM) certification \cite{gregory2023cism} and is appraised across six different metrics to gauge the quality of response readability.

The remaining of the paper is structured as follows: Section II reviews relevant literature, summarizing findings from related studies. Section III details the system's architecture and development process. Section IV presents the results and findings of our research. Finally, in Section V, we provide summaries, draw conclusions, and offer suggestions for future research directions.

\section{Background}
Adopting efficient cybersecurity strategies is challenging for small and medium-sized enterprises (SMEs) due to constraints such as limited budgets, a lack of skilled personnel, and insufficient time allocated to cybersecurity planning \cite{sme-survey}. Table \ref{tab:soc-cost} illustrates the typical expenses, totaling \$1,635,000, required to sustain a medium-sized Security Operations Center (SOC) team, accounting for both personnel and infrastructure. Effectively, this cost analysis sheds light on the challenges faced by smaller organizations and emphasizes the importance of accessible and cost-efficient cybersecurity solutions. 
The remainder of the section explores three key enabling technology areas contributing to the advancements in cybersecurity operations. 

\begin{table}[bp]
\caption{An estimate of the cost of a medium size SOC team \cite{sharkstriker-article}}
\begin{center}
\begin{threeparttable}
\centering
\begin{tabular}{|C{0.3\linewidth}|C{0.1\linewidth}|C{0.2\linewidth}|C{0.2\linewidth}|}
\hline \textbf{Role} & \textbf{Count} & \textbf{Yearly Cost} & \textbf{Total Cost}\\
\hline  Tier 1 Analyst & 5 & 80,000 \$ & 400,000 \$ \\
\hline  Tier 2 Analyst & 3 & 100,000 \$ & 300,000 \$ \\
\hline Tier 3 Analyst & 2 & 120,000 \$ & 240,000 \$ \\
\hline Malware Engineer  & 1 & 120,000 \$ & 120,000 \$ \\
\hline Forensic specialist & 1 & 130,000 \$ & 130,000 \$ \\
\hline SOC manager & 1 & 140,000 \$ & 140,000 \$ \\
\hline skill development & 13 & 5,000 \$ & 65,000 \$ \\
\hline Products/Infrastructure\tnote{*} & 1 & 240,000 \$ & 240,000 \$ \\
\hline \multicolumn{3}{|C{0.6\linewidth}|}{\textbf{Total cost for People / Infrastructure}} & 1,635,000 \$ \\
\hline
\end{tabular}
\begin{tablenotes}
\item[*] SIEM, EDR, Server Hardware, Forensic Software, and VAPT Tools
\end{tablenotes}
\end{threeparttable}
\label{tab:soc-cost}
\end{center}
\end{table}


\subsubsection{Network anomaly detection}
The purpose of an anomaly detection mechanism is to analyze, understand and characterize network traffic behavior, as well as to identify or classify the abnormal traffic instances such as malicious attempts from normal instances. Thus, from a machine learning perspective, the anomaly detection problem is a classification problem \cite{net-survey}. Over the years, detection systems have experienced considerable evolution, resulting in the development of diverse approaches and deployment methods, including those in fifth Generation (5G) communication networks and decentralised architectures.

Several Machine Learning (ML) techniques have been extensively applied in the domain of network anomaly detection, encompassing both supervised and unsupervised algorithms. Yihunie et al. \cite{rfc-ids} reviewed five representative algorithms: Stochastic Gradient Descent, Random Forests, Logistic Regression, Support Vector Machine, and Sequential Model, applying them to the NSL-KDD dataset. The empirical results from their study indicated that the Random Forest Classifier surpassed the other examined classifiers in terms of performance. 

Eltanbouly et al. \cite{hybrid-rfc-ids} introducing a hybrid system that combines the Random Forest and K-means algorithms. The proposed system is bifurcated into two distinct phases. The initial phase, known as the online phase, focuses on misuse detection by leveraging the Random Forest algorithm, followed by the offline phase, which categorizes random attacks through the use of the weighted K-means algorithm. Similarly, Zhao et al. \cite {10369705} proposed a Multi-Task Deep Neural Network in Federated Learning (MT-DNN-FL) to simultaneously detect network anomalies, recognize VPN (Tor) traffic, and classify traffic, while preserving data confidentiality. Experimental results on representative datasets demonstrated superior performance in detection and classification compared to several baseline methods in centralized training architecture.

Preuveneers et al. \cite{app8122663} proposed a blockchain-based federated learning method, allowing for the auditing of model updates without centralizing training data, thus providing enhanced transparency and accountability in detecting malicious behavior. The experiments show that while integrating blockchain increases complexity, the impact on performance is minimal (varying between 5 and 15\%), and the method is adaptable to more sophisticated neural network architectures and diverse use cases. Noticeably, the adoption of federated self-learning for anomaly detection and theat hunting is a rising trend in IoT devices, focusing on enhancing detection accuracy while prioritizing data privacy \cite{sheikhi2023cyber, 9424138, 8884802}.

Lately, the widespread adoption of 5G networks has increased the demand for the development of automated Intrusion Detection Systems, leading to a boost in specialized research in this domain \cite{porambage2021roadmap}. Sheikhi et al. \cite{5g-ids} focused on employing a federated learning-based method to identify DDoS attacks on the GTP protocol within a 5G core network. This approach capitalizes on the collective intelligence of various devices to proficiently and confidentially recognize DDoS attacks. While ML models exhibit strong performance in controlled settings, their efficacy in real-world environments is perceived as a significant barrier to their adoption \cite{cavallaro2023machine}. In the following section, we will discuss various explainability methods that have been proposed as potential solutions to overcome the barriers associated with the real-world performance of ML models.

\subsubsection{Explainable AI}
Explainable artificial intelligence (XAI) refers to advanced techniques that aim to make the results of machine learning  models understandable and more transparent to users. These techniques allow a practical deployment of ML models as they provide methods to ensure that the trained model is trustworthy by detecting biases in the model or in the corresponding training data, and increasing transparency in the predictions of the models by providing explanations for which input features had the most impact in the output of the model. For many critical applications in defense, medicine, finance, and law, explanations are essential for users to understand, trust, and effectively manage these new, artificially intelligent partners \cite{xai1}.

Recent research in Explainable Artificial Intelligence (XAI) has been actively applied to cybersecurity, particularly in specialized use cases like intrusion detection and malware identification. \cite{cybertrust}. Nguyen et al. develop GEE \cite{gee-xai}, a framework for detecting and explaining anomalies in network flow traffic. GEE comprises of two components. The first one consists of an unsupervised Variational Autoencoder (VAE) model for detecting network anomalies. The second one is a gradient-based fingerprinting technique for explaining the detected anomalies in the VAE. The evaluation shows that their approach is effective in detecting different anomalies as well as identifying fingerprints that are good representations of these attacks.

Han et al. developed the DeepAID framework \cite{deepai-xai} to interpret unsupervised DL-based anomaly detection systems for cybersecurity. The approach helps security analysts understand why a certain sample is considered anomalous by searching the difference between the anomaly and a normal reference data point. Additionally, they propose a model distiller that serves as an extension to the black-box DL model using a simpler and easier to understand finite-state machine model that allows analysts to get involved in the model decision-making process. While XAI holds promise in enhancing the adoption of ML models within existing cybersecurity frameworks, there are still several challenges and considerations to be address.

Nyre-Yu et al. \cite{nyre2022explainable} conducted a pilot study within an operational environment to assess an Explainable Artificial Intelligence (XAI) tool, focusing on insights gleaned from real-time interactions between cybersecurity analysts and XAI. The initial findings disclosed that, contrary to the goal of fostering trust and improving efficiency through XAI tools, their actual deployment was restricted and failed to considerably improve the accuracy of the decisions made by the analysts. Similarly, a recent systematic review \cite{nadeem2023sok} highlighted that research on XAI for cybersecurity shows that many XAI applications are crafted without a thorough understanding of their integration into analyst workflows. Moreover, security literature frequently fails to differentiate between diverse use cases or clearly separate the roles of model users and designers, which could potentially lead to diminished adoption.

\subsubsection{Chatbots for security \& ChatGPT} 
Conversational agents, often referred to as chatbots, have gained attention for their role in supporting cybersecurity within businesses through sharing network and security information with non-technical staff \cite{chatbot-article}. 

The Security focused chatbot introduced in \cite{secbot} named SecBot demonstrates the role of a  conversational agent for the support of cybersecurity planning and management. SecBot applies concepts of neural networks and Natural Language Processing (NLP), to interact and extract information from a conversation to (a) identify cyberattacks, (b) indicate solutions and configurations, and (c) provide insightful information for the decision on cybersecurity investments and risks.

Another notable advancement in Natural Language Understanding (NLU) with proven success in cybersecurity is the development of Generative Pre-trained Transformers (GPT) language models. 

These models can operate as standalone tools; for example, the application of GPT in formulating cybersecurity policies, as demonstrated in McIntosh et al. \cite{gpt-cs-policy}, heps deter and mitigate the impact of ransomware attacks involving data exfiltration. The results of the study indicated that, in specific scenarios, policies generated by GPT could surpass those created by humans, especially when supplied with customized input prompts. Similarly,  Setianto et al. \cite{10eeee} developed a run-time system, GPT-2C, that utilizes a fine-tuned GPT-2 model to parse logs from a live Cowrie SSH honeypot effectively, achieving 89\% inference accuracy in parsing Unix commands with minimal execution latency.

Furthremore, research emphasis has been placed on exploring the potential threats posed by Large Language Models (LLMs) like OpenAI’s ChatGPT and Bard, particularly regarding their abilities to facilitate cyberattacks  \cite{nour2023survey, sebastian2023chatgpt}. Nonetheless, limited research has been conducted on the potential integration of LLMs into cyber hunting interfaces. Bringing together conversational agents and GPT offers new opportunities for delivering knowledge and insights to non-professionals. Preliminary research and trials from other domains provide insight into this potential. For instance, \cite{financial-knowledge-gpt} examines the viability of leveraging Explainable AI (XAI) and language models like ChatGPT to transform how financial knowledge is conveyed to those outside the financial sector. The findings suggest that ChatGPT holds significant promise in demystifying intricate financial principles for a broader audience.

Finally, prompt engineering \cite{promptengineering}, an emerging research area, is showing promise as a potential solution for some of the shortcomings of LLMs since simple and effective prompts have been proven to improve GPT-3’s reliability. With appropriate prompts, GPT-3 is more reliable than smaller-scale supervised models on all these facets: generalizability, social biases, calibration, and factuality \cite{promptgpt}. However, evaluating the performance of Large Language Models across diverse domains presents an ongoing challenge. Several research studies have examined ChatGPT's performance in diverse fields. Gilson et al. \cite{gpt-med} investigated its performance on the United States Medical Licensing Examination, highlighting the ability of ChatGPT to provide logical and informative context in the majority of its responses. Notably, when ChatGPT delivers accurate and up-to-date information, users report higher satisfaction levels and make more informed decisions \cite{gpt-performance}. Therefore, understanding and quantifying the accuracy of the data provided in the cybersecurity domain is of high importance.

\section{ARCHITECTURE}

\begin{figure}[bp]
\centerline{\includegraphics[width=\linewidth]{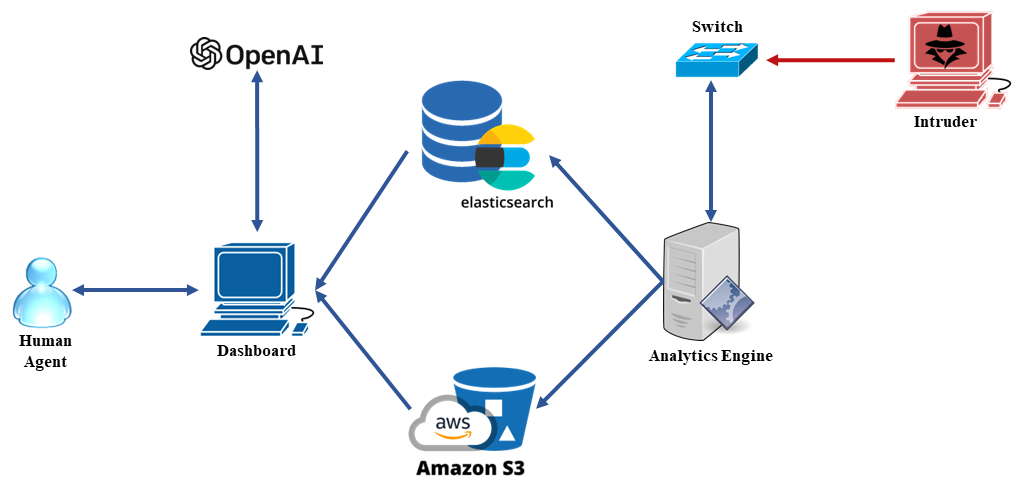}}
\caption{High level diagram of dashboard integration.}
\label{diagram}
\end{figure}

\begin{figure*}[htbp]
\centerline{\includegraphics[width=\linewidth]{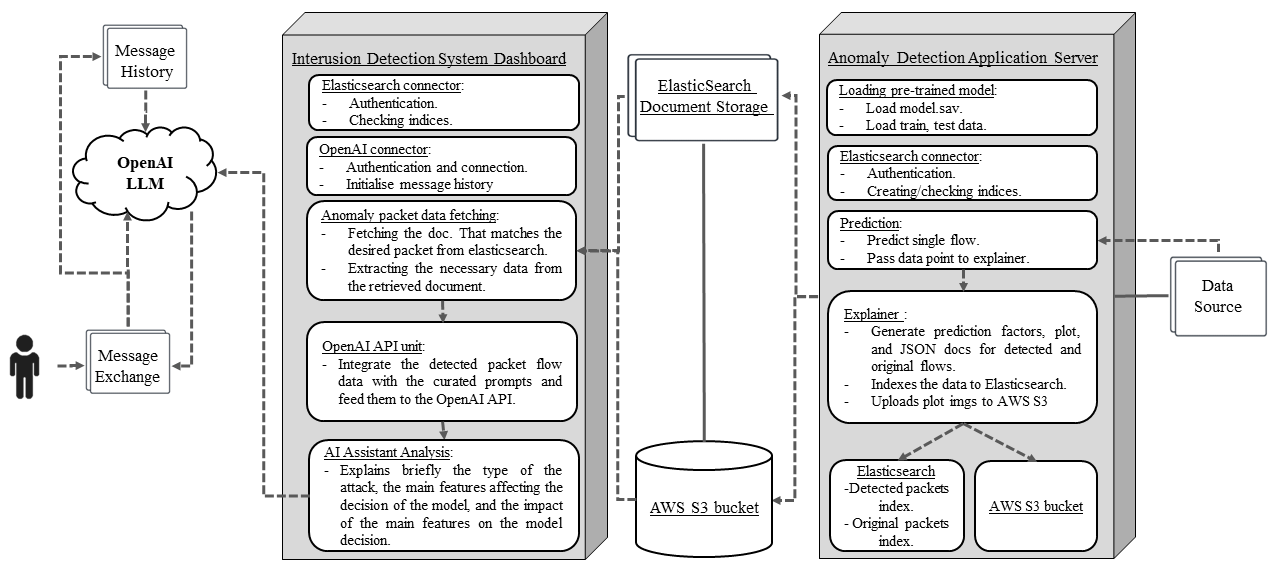}}
\caption{System Components Diagram}
\label{component-diagram}
\end{figure*}

Our system is organised into three distinct layers, each designed to fulfill specific functions and ensure optimal performance, as illustrated in Figure \ref{diagram}. Below, we outline the elements and functions of each layer, clarifying their roles and interactions within the overall system architecture.

\begin{enumerate}
\item Analytics engine: This is a powerhouse layer responsible for performing the network packet analysis, examines network data, identifying and processing anomalies and irregularities within the network flow.

\item Data Storage: We leverage Elasticsearch as our primary document storage, prized for its near real-time search capabilities, scalability, and reliability. It houses all detected network anomalies and the corresponding original flow data. For storing plots and images, Amazon S3 buckets are our go-to, guaranteeing security and accessibility.

\item User Interface (UI): The dashboard UI, constructed with Gradio, is the interactive front-end of our system, presenting the analytic engine’s outcomes to human analysts in a user-friendly manner. It is integrated with OpenAI's Language Model API, facilitating seamless interactions between analysts and the system for ongoing discussions and analysis.

\end{enumerate}

We chose this configuration due to its modular design, placing a premium on the segregation of components. This allows each layer to be developed, maintained, and enhanced independently, providing flexibility and promoting efficient scalability. The autonomous nature of each module ensures streamlined adaptability to evolving requirements, reinforcing the robustness and responsiveness of the system.

\subsection{Component Diagram}

In this section, we narrow our focus to provide more detailed insights into the architecture of our system, engaging in a concise discussion on each primary component. Figure \ref{component-diagram} depicts a comprehensive illustration of the diverse components comprising our system, which include i) the Anomaly Detection Application Server and ii) the Intrusion Detection System Dashboard.

\subsubsection{Anomaly Detection Application server} 
The Anomaly Detection Application Server serves as the central orchestrator of the entire anomaly detection process. This component integrates several sub-modules to facilitate the efficient and accurate identification of anomalous network behavior.

\begin{enumerate}

\item \textbf{ML Model Loader:} The First step in our anomaly detection pipeline is the loading of our pre-trained machine learning model. The model was extensively trained on the KDD99 Dataset \cite{kdd99} which is a widely used benchmark dataset in the field of intrusion detection and network security. We use the model to assess incoming data points against learned patterns. We also load the data into our engine which can be used to by explainability frameworks (SHAP and LIME) to provide interpretable explanations for predictions.

\item \textbf{Elasticsearch Connector:} Enabling seamless communication with Elasticsearch, the connector module handles authentication and index management, ensuring secure access to the Elasticsearch cluster. This components handles establishing a secure connection, creating and verifying indices, eventually enabling efficient storage of information pertaining to detected and original packets.

\item \textbf{Prediction}: The prediction component analyzes individual network flows to determine the presence of anomalies.

\item \textbf{Explainer}: This component generates prediction factors, plots, and JSON documents. The explainer indexes the generated data into Elasticsearch, constructing a structured repository that facilitates efficient querying and exploration of detected and original network flows. To augment the interpretability of our findings, the explainer component uploads plot images to an AWS S3 bucket. These plots enhance our understanding of the model dynamics.

\item \textbf{Elasticsearch}: Elasticsearch plays a pivotal role in storing and organizing information. Our system leverages Elasticsearch to manage both the "Detected Packets Index" and the "Original Packets Index," optimizing data accessibility and analysis.

\item \textbf{AWS S3 Bucket:} Serving as a centralized repository for our visual resources housing the uploaded plots. 

\end{enumerate}

\subsubsection{Intrusion Detection System Dashboard}
Our IDS Dashboard enhances trust in the AI system by revealing model insights and contributing features. It lets users inspect original data packets for manual anomaly inspection. AI-generated explanations further clarify predictions and suggest appropriate course of action. The dashboard also facilitates interactive discussions with the AI assistant, offering custom insights into the model's operations. This component integrates several sub-modules to facilitate its work in presenting explainability and fairness to the end user.

\begin{enumerate}
\item \textbf{OpenAI Connector:} The OpenAI Connector is mainly used for authentication with the OpenAI API and also initiaties the prerecorded message history. It also keeps track of the user conversations.
\item \textbf{Anomaly Packet Data Fetching:} It looks through all the documents in Elasticsearch index and extracts all the important information we need from that document.
\item \textbf{OpenAI API Unit:} Integrates the detected packet flow data with the curated fine-tuning prompts and feed them to the OpenAI API.
\item \textbf{AI Assistant Analysis:} The AI Assistant receives all the data preserved in the document, accompanied by the refined prompts from the unit, and it generates a comprehensive analysis for the human agent. This analysis not only reveals the details but also facilitates interactive communication directly with the human agent, enabling a seamless exchange of information and insights.

\end{enumerate}

\subsection{Elasticsearch data schemas}


For telemetry storage, we principally utilize two Elasticsearch indices: detected-packets, original-packets. The "detected-packets" index serves as a structured repository for crucial information related to detected network anomalies. This schema is designed to store the attributes extracted using the two explainability frameworks SHAP and LIME, each contributing to the comprehensive understanding of identified anomalies.

The schema for the "detected-packets" index is designed to align with the objective of offering insights into the workings of the anomaly detection model, ensuring transparency and a deeper understanding of its mechanisms. It offers a structured means of storing textual predictions, influential factors, and other data, enhancing the dashboard's capabilities to explain anomalies effectively. Table \ref{tab:detected-packets-index} provides an overview of the "Detected Packets" index, detailing each label and its corresponding description to ensure clarity and a comprehensive understanding of the data structure. The table emphasizes key attributes pivotal for anomaly detection and subsequent analysis.

The "original-packets" index requires a more complex and dynamic mapping. An in-depth review of each field in the document is superfluous since the KDD99 dataset description \cite{kdd99} already offers a comprehensive coverage. The schema for the original packets is crucial, aiding anomaly detection through manual reviews and enhancing trust between the human agent and the Intrusion Detection System (IDS).

\subsection{User Experience \& Use Cases}

Collectively, all the elements of the IDS Dashboard we previously discussed assist the user in making well-informed decisions regarding network anomalies by incorporating explanatory visualizations, manual inspection, AI-generated explanations, and interactive conversations. The system is intended to be versatile and can aid the user in multiple scenarios. 

First, the Detection Engine offers updated threat identification and classification of incidents, swiftly classifying network threats and allowing security analysts to make informed responses through insightful visualizations and AI-generated explanations available on the Dashboard. Second, the dashboard enables ML model Interpretability, serving as a crucial tool for developers and data scientists to comprehend model functionality and its shortcomings, aiding the creation of more advanced models. Third, addressing the prevalent concern of soaring security operation costs, our dashboard promotes collaborative analysis and reporting by providing a user-friendly interface suitable for individuals with varied security expertise and equipping users with essential tools, thereby optimizing operational effectiveness and cost efficiency.

\begin{table}[tp]
\caption{Detected Packets Index}
\begin{center}
\begin{tabular}{|C{0.2\linewidth}|C{0.7\linewidth}|}
\hline \textbf{Label} & \textbf{Description} \\
\hline  \_id & Represents a unique identifier for each detected packet entry. \\
\hline  Prediction & stores the prediction outcome associated with the detected packet wther normal or malicious and also the specific genre of the attack. \\
\hline Factors &  Captures the contributing factors that influenced the prediction outcome and the key attributes that led to the anomaly's detection. We store textual descriptions of these factors, to be later translated into visual plots and also fed into the chatbot to provide contextual answers.  \\
\hline Exp-img  & Stores references to the plots used in presenting the prediction process, enhancing the comprehensibility of the anomaly detection process. Exp-img represents the local explanation plot provided by LIME tabular explainer. \\
\hline Shap-img & Shap-img represents the  top features contributing to the predicted anomaly class using SHAP framework. \\
\hline Original-data & Denotes the number of the detected packet which can be used as a foreign key to access the complete information regarding the detected packets. \\
\hline
\end{tabular}
\newline
\label{tab:detected-packets-index}
\end{center}
\end{table}

\begin{figure*}[tp]
\centerline{\includegraphics[width=\linewidth]{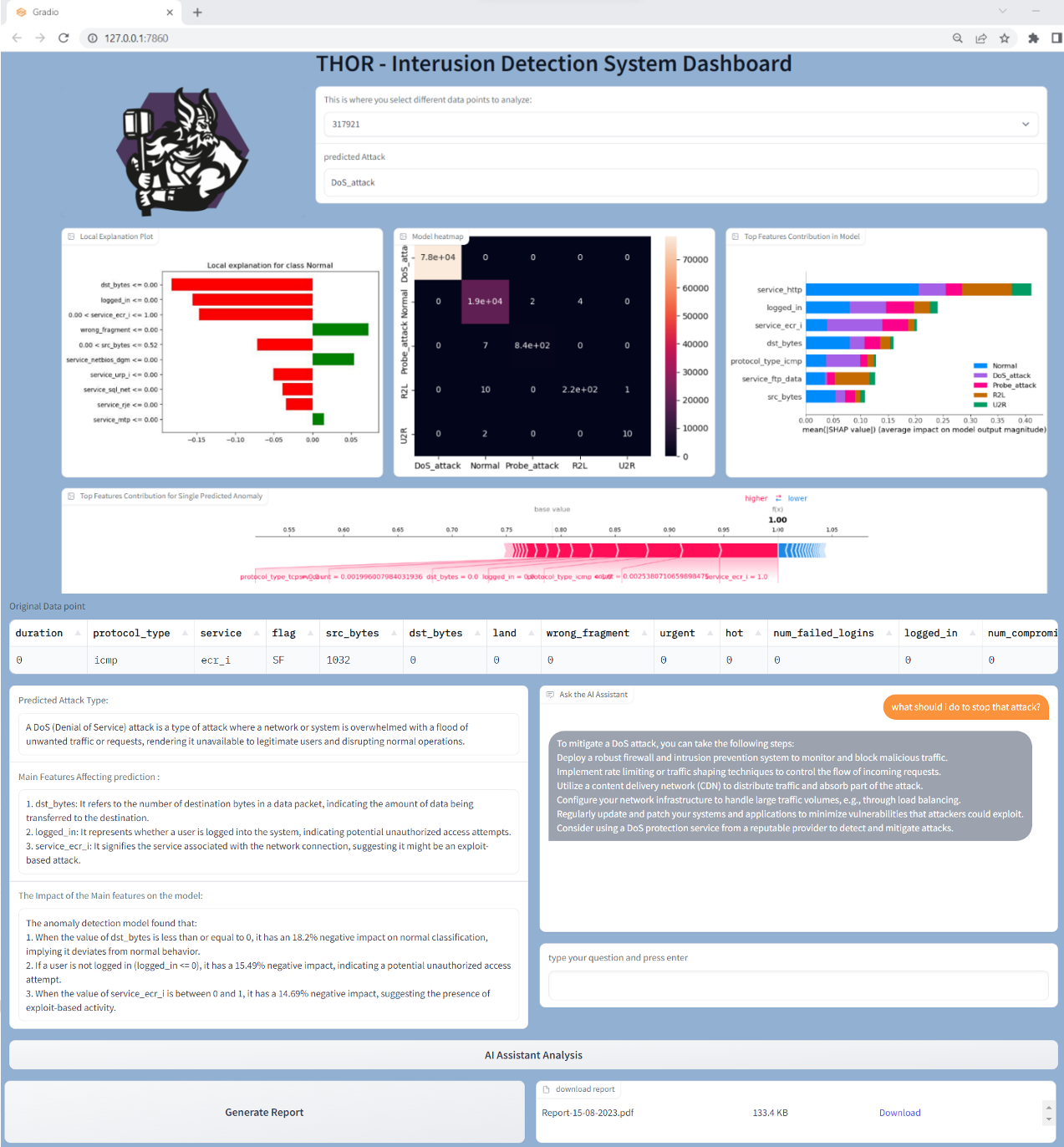}}
\caption{Sections of Detection and Explainability in the Dashboard.}
\label{detection-section}
\end{figure*}

\section{EVALUATION AND RESULTS}

\subsection{Prototype Functionality}
The dashboard shown in Figure \ref{detection-section} features a neatly organized layout that seamlessly integrates the multiple components of the HuntGPT. In the upper section, the dashboard provides a visual representation of the ML model inner workings. Moving down,  the middle area presents the original data for manual inspection when needed.  Lastly, the lower part is the AI assistant that provides context and explainability to the end user and also enables on going conversations. The system user can download a complete report of the incident including all the graphs an data regarding the incident using the generate report function. This setup ensures user-friendliness while uniting various functions.

\subsection{Response Quality Analysis}
In this section, we examine how well our prototype functions in explaining detected anomalies and assisting users via its chatbot feature. Assessing the responses of the chatbot, a common practice with conversational agents, posed distinctive challenges in quantifying the requisite metrics for performance appraisal. Our analysis can be segmented into two key components: i) Technical Knowledge in Cybersecurity and ii) Response Evaluation. 

We first examine whether ChatGPT (i.e., GPT-3.5 Turbo) possesses the requisite technical knowledge in the field of cybersecurity to effectively assist the user. A critical aspect of this analysis involves comparing ChatGPT's knowledge to that of a certified IT professional. Following this knowledge assessment, we evaluate the responses provided by ChatGPT in terms of their quality and appropriateness. Special attention is paid to the level of difficulty in the answers since our goal is to provide knowledge to users with minimal cybersecurity experience.

\subsubsection{Technical Knowledge in Cybersecurity}
In this assessment, our primary focus is to measure the accuracy and precision of answers generated by GPT-3.5-turbo, the model currently powering the HuntGPT system. To evaluate the model's performance, we conducted tests using a set of standardized certification exams in the field of cybersecurity. The detailed list of the standardized exams used for our assessment is presented below:

    \begin{enumerate}
        \item CISM Certified Information Security Manager Practice Exams:  The updated self-study guide Written by Peter H. Gregory featuring hundreds of practice exam questions that match those on the live test \cite{gregory2023cism}.
        \item ISACA official CISM practice Quiz. A free practice quiz including questions from ISACA's test prep solutions that are the same level of difficulty of ISACA's official CISM exam \cite{isaca}.
        \item ISACA official cybersecurity fundamentals practice quiz: A practice quiz including questions from ISACA's test prep solutions that are the same level of difficulty of ISACA's official Cybersecurity Fundamentals exam \cite{isaca}.
    \end{enumerate}

\begin{table}[bp]
\caption{Comparison of CISM and Cybersecurity Exam Preparations: GPT-3.5 Success Rates}
\centering
\resizebox{\columnwidth}{!}{%
\begin{tabular}{|p{0.4\linewidth}|c|c|}
\hline
\multicolumn{1}{|c|}{\textbf{Exam}} & \multicolumn{1}{c|}{\textbf{No. of Questions}} & \multicolumn{1}{c|}{\textbf{GPT-3.5 turbo Success Rate}} \\
\hline
CISM Certified Information Security Manager Practice Exams \cite{gregory2023cism} & 40 & 82.5\% \\
\hline
ISACA official CISM practice Quiz \cite{isaca} & 10 & 80\% \\
\hline
ISACA official cybersecurity fundamentals practice quiz \cite{isaca} & 25 & 72\% \\
\hline
\end{tabular}%
}
\label{tab:exam-gpt-success}
\end{table}

Based on the results presented in Table \ref{tab:exam-gpt-success}, GPT-3.5-turbo demonstrates substantial proficiency in cybersecurity knowledge, with success rates ranging from 72\% to 82.5\% across varied and reputable exams. However, there is room for improvement, particularly in the ISACA official cybersecurity fundamentals practice quiz where the model achieved a lower success rate of 72\%. The challenging CISM certification demands profound knowledge and experience in risk management, illustrated by the fact that only 50-60\% of first-time examinees attain success \cite{cism-article}. Nevertheless, preliminary results from our evaluation are highly promising and indicate that the model possesses the capacity to provide well-informed security decisions.

\begin{table}[bp]
\caption{evaluating generated text readability}
\begin{center}

\begin{tabular}{|C{0.4\linewidth}|C{0.05\linewidth}|C{0.10\linewidth}|C{0.05\linewidth}|C{0.10\linewidth}|}
\hline \multicolumn{1}{|C{0.4\linewidth}|}{\multirow{2}{*}{\textbf{Readability Formula}}} & \multicolumn{2}{C{0.2\linewidth}}{\textbf{Generated Anomaly explanation}} & \multicolumn{2}{|C{0.2\linewidth}|}{\textbf{Chatbot Answers}}\\
\cline{2-5}   \multicolumn{1}{|C{0.45\linewidth}|}{}  & \textbf{Score} & \textbf{Grade level} & \textbf{Score} & \textbf{Grade level} \\
\hline  Flesch-Kincaid Grade Level & 15.7 & 16 & 14.9 & 15 \\
\hline  Flesch Reading Ease &  22.7 & graduate & 23.9 & graduate \\
\hline Dale Chall & 12.6 & graduate  & 12.11 & graduate\\
\hline Automated Readability Index & 16.3 & graduate & 15.9 & graduate   \\
\hline Coleman Liau Index & 15.3 & 15 & 16.3 & 16 \\
\hline Linsear Write & 17.4 & 17 & 15.8 & 16\\
\hline
\end{tabular}
\newline
\label{tab:text-readability}
\end{center}
\end{table}

\subsubsection{Response Evaluation}
We evaluated the responses provided by GPT-3.5-turbo in terms of their quality and appropriateness. Table \ref{tab:text-readability} presents the comprehensive findings of our study, where we evaluated 20 different generated explanations and 20 chat logs. Each chat log included approximately six questions regarding the detected anomaly. An example of a chat log is illustrated in Table \ref{tab:sample-chat}.

We evaluated the responses using six unique readability formulas from the Python \textit{py-readability-metrics} library \cite{pyreadabilitymetrics}, each applicable to different fields due to their distinct characteristics. The Flesch-Kincaid Grade Level, adopted by the U.S. Army, is used for assessing the complexity of technical manuals. The Flesch Reading Ease serves as a standard test for the U.S. Department of Defense, employed to determine the readability of its documents and forms. The Dale Chall readability formula is widely adopted in schools and educational institutes to assess the difficulty level of texts. The Automated Readability Index was initially designed for real-time monitoring of readability on electric typewriters. The Coleman Liau Index has been specifically used to evaluate the readability of medical documents. Finally, the Linsear Write was purportedly developed for the United States Air Force to assess the readability of their technical manuals.

The readability results in Table \ref{tab:text-readability} for both AI explainability and Chatbot logs were evaluated by most formulas as being at the graduate level or equivalent, revealing the presence of complex linguistic structures in the generated texts. While chat conversations can display different readability levels depending on question types and user-required detail, the generated text is usually readable, typically necessitating only a basic, non-specialized college education level for comprehension.

\begin{table}[tp]
\scriptsize
\caption{Examples of submitted queries to HuntGPT and the related actionable responses}
\begin{center}
\begin{tabular}{|C{0.2\linewidth}|C{0.7\linewidth}|}
\hline \textbf{Query} & \textbf{Chatbot Response} \\
\hline How can I prevent such attack & To prevent DoS attacks, you can implement measures like traffic monitoring, firewalls, load balancers, and rate limiting to detect and mitigate abnormal traffic patterns. Ensuring network redundancy and having a robust incident response plan can also help minimize the impact. \\
\hline What is a firewall and how do I implement it with some good firewall examples  & A firewall is a network security device that filters incoming and outgoing network traffic based on predetermined security rules. Examples of good firewalls include:
Cisco ASA (Adaptive Security Appliance), Palo Alto Networks Next-Generation Firewalls, Fortinet FortiGate Firewalls, Check Point Firewall. To implement a firewall, configure rule sets to allow or block specific types of network traffic and define security policies to secure your network from unauthorized access. \\
\hline Are there any free ones? &  Yes, there are free firewall solutions available. Some popular free firewall options include:
ZoneAlarm: ZoneAlarm offers a free version of their firewall software for personal use, providing basic firewall protection along with additional features like identity protection and anti-phishing.
Windows Firewall: If you are using a Windows operating system, it comes with a built-in firewall called Windows Firewall. It provides basic inbound and outbound traffic filtering capabilities.
These free firewall solutions can offer varying levels of protection and features. \\
\hline  How can I install zonealarm & To install ZoneAlarm, follow these steps:
Visit the official ZoneAlarm website (www.zonealarm.com) using your web browser.
Go to the "Products" or "Downloads" section of the website. Look for the free version of ZoneAlarm Firewall.
\textbf{[proceeds to give detailed instrctions on downloading ZoneAlarm]}
\\
\hline What would be a good configuration for a small home network & To configure ZoneAlarm Firewall for a small home network:
Install and update ZoneAlarm.
Use medium or high-security mode.
Enable automatic program control.
Customize program permissions.
Enable Stealth Mode for added security.
Set up Network Zones.
Monitor network traffic.
Keep ZoneAlarm updated.  \\
\hline
\end{tabular}
\newline
\label{tab:sample-chat}
\end{center}
\end{table}

\section{CONCLUSION AND FUTURE WORK}

In conclusion, the study illustrates the efficacy of integrating LLM-based conversational agents with Explainable AI (XAI) within intrusion detection systems. Our prototype, HuntGPT, combines the advanced capabilities of GPT-3.5-turbo with a user-friendly dashboard and adeptly elucidates the latent details of detected anomalies. The model demonstrated substantial proficiency in cybersecurity knowledge, as evident from its success rates ranging between 72\% and 82.5\% on various reputable standardized certification exams in cybersecurity. However, these results also highlight areas for enhancement, mainly focusing on improving proficiency in fundamental cybersecurity concepts where the model achieved a lower success rate.

The extensive readability analysis of generated responses indicates that the content produced is generally comprehensible for individuals with a basic college education level, fostering user understanding and interaction. The implemented conversational agent effectively generates actionable responses and promotes user engagement by communicating complex concepts, thereby making a substantial contribution to enhancing user comprehension in cybersecurity.

This research contributes valuable insights into integrating advanced AI models within interactive, user-focused applications in cybersecurity. The achieved success rates and the readability of generated responses emphasize the potential of implementing such integrated models in real-world applications, providing a solid foundation for developing more sophisticated, explainable, interpretable,  actionable, and  user-friendly cybersecurity solutions. This study serves as a stepping stone for further research and development in creating LLM-driven security tools that integrate XAI  in response to the evolving landscape of cybersecurity threats and user needs.

In our future work, we aim to advance in two simultaneous directions, which are crucial for meeting the real-time detection and response requirements. Our first approach involves refining our Machine Learning model and incorporating real-time detection capabilities by adapting existing frameworks, thereby enabling immediate and proactive responses to threats. Concurrently, we focus on embedding actionable AI within the chatbot, enabling it to issue active commands directly to the Security Information and Event Management (SIEM) component, facilitating instantaneous and intelligent responses to security events.

\section*{Acknowledgment}
This work was funded by the European Commission grants IDUNN (grant no. 101021911) and the Academy of Finland 6Genesis Flagship program (grant no. 318927).
\printbibliography
\end{document}